\documentclass[10pt,twocolumn,twoside]{IEEEtran}

\usepackage{graphicx}
\usepackage{cite}
\usepackage[cmex10]{amsmath}
\usepackage{amssymb}

\usepackage{tikz}
\usepackage{pgfplots}
\pgfplotsset{compat=1.3}
\usetikzlibrary{arrows,snakes,shapes}

\DeclareMathAlphabet{\mathbit}{OML}{cmr}{bx}{it}

\DeclareMathOperator{\Q}{Q}

\DeclareMathOperator{\E}{E}
\DeclareMathOperator{\T}{T}

\DeclareMathOperator{\Probability}{Pr}
\DeclareMathOperator{\Diag}{diag}

\renewcommand\vec[1]{\operatorname{vec}\left(#1\right)}
\renewcommand\arcsin[1]{\operatorname{arcsin}\left(#1\right)}

\DeclareMathOperator{\fieldR}{\mathbb{R}}

\newcommand{\diag}[1]{\Diag{\left(#1\right)}}

\newcommand{\sign}[1]{\Sign{\left(#1\right)}}

\newcommand{\ve}[1]{\boldsymbol{#1}}

\newcommand{\exdi}[2]{\E_{#1} \left[#2\right]}

\renewcommand{\exp}[1]{\operatorname{exp}\left(#1\right)}

\newcommand{\Prob}[1]{\Probability\left\{#1\right\}}

\newcommand{\qfunc}[1]{\Q \left(#1\right)}
\newcommand{\qfuncinv}[1]{\Q^{-1} \left(#1\right)}

\newcommand\Sign{\operatorname{sign}}

\newcommand{\expbrace}[1]{\operatorname{exp} \left(#1\right)}

\title{Asymptotic Signal Detection Rates\\with $1$-bit Array Measurements}

\author{Manuel S. Stein\thanks{This work was supported by the German Academic Exchange Service (DAAD) with funds from the German Federal Ministry of Education and Research (BMBF) and the People Program (Marie Sk{\l}odowska-Curie Actions) of the European Union's Seventh Framework Program (FP7) under REA grant agreement no. 605728 (P.R.I.M.E. - Postdoctoral Researchers International Mobility Experience).}
\thanks{M. S. Stein is with the Chair for Stochastics, Universit\"at Bayreuth, Germany (e-mail: manuel.stein@uni-bayreuth.de).}
}

\begin{document}
\maketitle
\begin{abstract}
This work considers detecting the presence of a band-limited random radio source using an antenna array featuring a low-complexity digitization process with single-bit output resolution. In contrast to high-resolution analog-to-digital conversion, such a direct transformation of the analog radio measurements to a binary representation can be implemented hardware and energy-efficient. However, the probabilistic model of the binary receive data becomes challenging. Therefore, we first consider the Neyman-Pearson test within generic exponential families and derive the associated analytic detection rate expressions. Then we use a specific replacement model for the binary likelihood and study the achievable detection performance with $1$-bit radio array measurements. As an application, we explore the capability of a low-complexity GPS spectrum monitoring system with different numbers of antennas and different observation intervals. Results show that with a moderate amount of binary sensors it is possible to reliably perform the monitoring task.
\end{abstract}
\section{Introduction}\label{sec:intro}
Since, in 1965, Gordon E. Moore predicted a doubling in computational capability every two years, chip companies have kept pace with this prognosis and set the foundation for digital systems which today allow processing high-rate radio measurements by sophisticated algorithms. In conjunction with wireless sensor arrays, this results in advanced signal processing capabilities, see, e.g. \cite{Krim96}. Unfortunately, in the last decades, the advances regarding the analog circuits forming the radio front-end were much slower. Therefore, in the advent of the Internet of things (IoT), where small and cheap objects are supposed to feature radio interfaces, cost and power consumption of wireless sensors are becoming an issue. In particular, analog-to-digital conversion \cite{Walden99,MurmannSurvey} shows to set constraints on the digitization rate and the number of antenna elements under strict hardware and power budgets. 

In this context, we consider detection of a band-limited source with unknown random structure by a sensor array providing single-bit radio measurements. We formulate the processing task as a binary hypothesis test regarding exponential family models and derive expressions for the asymptotic detection rates. The results are used to determine the design of a low-complexity GPS spectrum monitoring system.

Note that detection with quantized signals has found attention in distributed decision making \cite{Sandell81} where data is collected through sensors at different locations and quantization is used to diminish the communication overhead between sensors and terminal node \cite{Fang13,Ciuonzo13,Zayyani16}. Optimum quantization for detection is the focus of \cite{Kassam77,Poor77,Aazhang84,Hashlamoun96}, while \cite{Willett95} considers the detection performance degradation due to hard-limiting. For discussions on symbol detection for quantized communication see, e.g., \cite{Foschini76,Dabeer03,Mezghani08,Wang15, Choi16,Jacobsson17}. Detection for cognitive radio with single-antenna $1$-bit receivers is considered in \cite{Ali16} while array processing with $1$-bit measurements is analyzed in \cite{BarShalom02,SteinWSA16,Liu17}.
\section{Problem Formulation}\label{sec:problem:form}
We consider a receive situation where a narrow-band random wireless signal is impinging on an array of $S$ sensors,
\begin{align}\label{receive:model:unquantized}
\ve{y}=\gamma\ve{A}\ve{x}+\ve{\eta}.
\end{align}
Each sensor features two outputs (in-phase and quadrature), such that the receive vector $\ve{y}\in\ve{\mathcal{Y}}=\fieldR^M$, $M=2S$, can be decomposed $\ve{y}=\begin{bmatrix}\ve{y}^{\T}_\text{I} &\ve{y}^{\T}_\text{Q} \end{bmatrix}^{\T}$ with $\ve{y}_\text{I},\ve{y}_\text{Q}\in\fieldR^S$. Likewise, the random source $\ve{x}=\begin{bmatrix}x_\text{I} &x_\text{Q} \end{bmatrix}^{\T}\in\fieldR^2$ consists of two zero-mean signal components with covariance matrix
\begin{align}
\ve{R}_{\ve{x}}=\exdi{\ve{x}}{\ve{x}\ve{x}^{\T}}&=\ve{I},
\end{align}
where $\exdi{\ve{u}}{\cdot}$ denotes the expectation concerning the probability distribution $p(\ve{u})$ and $\ve{I}$ the identity matrix. The steering matrix $\ve{A}=\begin{bmatrix}\ve{A}^{\T}_\text{I} &\ve{A}^{\T}_\text{Q} \end{bmatrix}^{\T}\in\fieldR^{M \times 2},$ with $\ve{A}_\text{I}, \ve{A}_\text{Q}\in\fieldR^{S \times 2}$ models a uniform linear sensor array response (half carrier-wavelength inter-element distance) for a narrow-band signal arriving from direction $\zeta\in\fieldR$, such that
\begin{align}
\ve{A}_\text{I}=\begin{bmatrix}
\cos{\big(0\big)} &\sin{\big(0\big)}\\ 
\cos{\big(\pi\sin{(\zeta)}\big)} &\sin{\big(\pi\sin{(\zeta)}\big)}\\ 
\vdots &\vdots\\ 
\cos{\big((S-1)\pi\sin{(\zeta)}\big)} &\sin{\big((S-1)\pi\sin{(\zeta)}\big)}
\end{bmatrix}
\end{align}
and
\begin{align}
\ve{A}_\text{Q}=\begin{bmatrix}
-\sin{\big(0\big)} &\cos{\big(0\big)}\\ 
-\sin{\big(\pi\sin{(\zeta)}\big)} &\cos{\big(\pi\sin{(\zeta)}\big)} \\ 
\vdots &\vdots\\ 
-\sin{\big((S-1)\pi\sin{(\zeta)}\big)} &\cos{\big((S-1)\pi\sin{(\zeta)}\big)}
\end{bmatrix}.
\end{align}
The parameter $\gamma\in\fieldR$ characterizes the source strength in relation to the additive zero-mean sensor noise $\ve{\eta}\in\fieldR^M$ with
\begin{align}
\ve{R}_{\ve{\eta}}=\exdi{\ve{\eta}}{\ve{\eta}\ve{\eta}^{\T}}=\ve{I}.
\end{align}
Due to the properties of the source and noise signals, the receive data \eqref{receive:model:unquantized} can be modeled by a Gaussian distribution
\begin{align}\label{multivariate:gauss}
\ve{y}\sim p_{\ve{y}}(\ve{y};\gamma)=\frac{\exp{-\frac{1}{2} \ve{y}^{\T} \ve{R}_{\ve{y}}^{-1}(\gamma) \ve{y}}}{ \sqrt{(2\pi)^{M} \det{(\ve{R}_{\ve{y}}(\gamma))}} },
\end{align}
with covariance matrix
\begin{align}
\ve{R}_{\ve{y}}(\gamma)&=\exdi{\ve{y};\gamma}{\ve{y}\ve{y}^{\T}}=\gamma^2\ve{A}\ve{A}^{\T}+\ve{I}.
\end{align}
Based on the likelihood \eqref{multivariate:gauss}, the problem of signal detection can be stated as the decision about which of the two models
\begin{align}
\mathcal{H}_0: \ve{y}\sim p_{\ve{y}}(\ve{y};\gamma_0),\quad\mathcal{H}_1: \ve{y}\sim p_{\ve{y}}(\ve{y};\gamma_1)
\end{align}
has generated the data ($K$ independent array snapshots)
\begin{align}\label{quantized:data}
\ve{Y}=\begin{bmatrix} \ve{y}_1 &\ve{y}_2 &\ldots &\ve{y}_K\end{bmatrix} \in \ve{\mathcal{Y}}^{K}.
\end{align}
Using model \eqref{receive:model:unquantized} in practical applications implies that a high-resolution analog-to-digital converter (ADC) is available for each output channel. Significant savings are possible when using a converter with $1$-bit output resolution. Such a receiver can be modeled by an ideal sampling device with infinite resolution followed by a hard-limiter
\begin{align}\label{system:model:sign}
\ve{z} = \sign{\ve{y}},
\end{align}
where $\sign{\ve{u}}$ is the element-wise signum function, i.e.,
\begin{align}
\left[\ve{z}\right]_i=
\begin{cases}
+1& \text{if } [\ve{y}]_i \geq 0\\
-1 & \text{if } [\ve{y}]_i < 0.
\end{cases}
\end{align}
Note, that \eqref{system:model:sign} characterizes a low-complexity digitization process without feedback and is, therefore, distinct from sigma-delta conversion \cite{Aziz96}, where a single fast comparator with feedback is used to mimic a high-resolution ADC.

Modeling the output of \eqref{system:model:sign} by its exact parametric probability distribution function, requires computing the integral
\begin{align}\label{likelihood:quantizer}
p_{\ve{z}}(\ve{z};\gamma)=\int_{\ve{\mathcal{Y}}(\ve{z})} p_{\ve{y}}(\ve{y};\gamma) {\mathrm d}\ve{y}
\end{align}
for all $2^M$ points in $\ve{\mathcal{Z}}=\mathbb{B}^{M}$, where $\ve{\mathcal{Y}}(\ve{z})$ characterizes the subset in $\ve{\mathcal{Y}}$ which by \eqref{system:model:sign} is transformed to $\ve{z}$. Additionally, \eqref{likelihood:quantizer} requires the orthant probability, which for $M>4$ is an open problem. As the multivariate Bernoulli model resulting from \eqref{system:model:sign} is part of the exponential family like \eqref{multivariate:gauss}, in the following we resort to discussing the considered processing task for generic data models within this broad class. Without exactly specifying the quantized likelihood \eqref{likelihood:quantizer}, this will allow us to analyze the asymptotically achievable detection rates with low-complexity $1$-bit array measurements.
\section{Decisions in the Exponential Family}\label{sec:format}
Consider the multivariate parametric exponential family 
\begin{align}\label{replacement:exp:family}
{p}_{\ve{z}}(\ve{z};\ve{\theta})=\exp{\ve{\beta}^{\T}(\ve{\theta}) \ve{\phi}(\ve{z}) - \lambda(\ve{\theta})+\kappa(\ve{z})},
\end{align}
where $\ve{\theta}\in\fieldR^{D}$ constitute its parameters, $\ve{\beta}(\ve{\theta}): \fieldR^{D} \to \fieldR^{L}$ the natural parameters, $\ve{\phi}(\ve{z}): \fieldR^{M} \to\fieldR^{L}$ the sufficient statistics, $\lambda(\ve{\theta}): \fieldR^{D} \to \fieldR$ the log-normalizer and $\kappa(\ve{z}): \fieldR^{M} \to\fieldR$ the carrier measure. Given a data set $\ve{Z}\in\ve{\mathcal{Z}}^{K}$ of the form \eqref{quantized:data}, the simple binary hypothesis test between
\begin{align}\label{bin:test:exp:fam}
\mathcal{H}_0: \ve{z}\sim {p}_{\ve{z}}(\ve{z};\ve{\theta}_0),\quad\mathcal{H}_1: \ve{z}\sim {p}_{\ve{z}}(\ve{z};\ve{\theta}_1)
\end{align}
is to be performed. To this end, we assign a critical region $\ve{\mathcal{C}}\subset \ve{\mathcal{Z}}^K$ and decide in favor of $\mathcal{H}_1$ if the observed data satisfies $\ve{Z}\in\ve{\mathcal{C}}$. The probability of erroneously deciding for $\mathcal{H}_1$ while $\mathcal{H}_0$ is the true data-generating model is calculated
\begin{align}\label{prob:fa}
P_{\text{FA}}=\int_{\ve{\mathcal{C}}} {p}_{\ve{Z}}(\ve{Z};\ve{\theta}_0) {\rm d}\ve{Z},
\end{align}
while the probability of correctly deciding for $\mathcal{H}_1$ is given by
\begin{align}\label{prob:de}
P_{\text{D}}=\int_{\ve{\mathcal{C}}} {p}_{\ve{Z}}(\ve{Z};\ve{\theta}_1) {\rm d}\ve{Z}.
\end{align}
Approaching the decision problem \eqref{bin:test:exp:fam} under the desired test size $P_{\text{FA}}$ and maximum $P_{\text{D}}$, the Neyman-Pearson theorem shows that it is optimum to use the likelihood ratio test \cite{Kay98}
\begin{align}\label{definition:LR:Neyman:Pearson}
L(\ve{Z})=\frac{{p}_{\ve{Z}}(\ve{Z};\ve{\theta}_1)}{{p}_{\ve{Z}}(\ve{Z};\ve{\theta}_0)}>\xi'
\end{align}
for the assignment of the critical region 
\begin{align}
\ve{\mathcal{C}}={\{\ve{Z}: L(\ve{Z})>\xi' \}}, 
\end{align}
while the decision threshold $\xi'$ is determined through
\begin{align}
P_{\text{FA}}=\int_{\{\ve{Z}: L(\ve{Z})>\xi' \}} {p}_{\ve{Z}}(\ve{Z};\ve{\theta}_0) {\rm d}\ve{Z}.
\end{align}
Based on the ratio \eqref{definition:LR:Neyman:Pearson}, a test statistic $T(\ve{Z}): \ve{\mathcal{Z}}^K \to \fieldR$ can be formulated such that the binary decision is performed by
\begin{align}\label{def:decision:process}
\text{decide } \begin{cases} \mathcal{H}_0 &\mbox{if}\quad T(\ve{Z})\leq \xi \\ 
\mathcal{H}_1 & \mbox{if}\quad T(\ve{Z}) > \xi \end{cases}.
\end{align}
To analyze the performance of \eqref{def:decision:process}, it is required to characterize the distribution of $T(\ve{Z})$ and evaluate \eqref{prob:fa} and \eqref{prob:de}.
As the data $\ve{Z}$ consists of $K$ independent samples, the test statistic can be factorized into a sum of independent components
\begin{align}
T(\ve{Z})&=\sum_{k=1}^{K} t(\ve{z}_k)
\end{align}
such that, by the central limit theorem, the test statistic in the large sample regime follows the normal distribution
\begin{align}\label{def:test:statistic:gaussian:approx}
p\big(T(\ve{Z})| \mathcal{H}_i\big)&\overset{a}{=}\frac{1}{\sqrt{2 \pi}\sigma_i} \expbrace{- \frac{(T(\ve{Z}) - \mu_i)^2}{2\sigma_i^2} },
\end{align}
where by $\overset{a}{=}$ we denote asymptotic equality. Through the mean and standard deviation of the test statistic
\begin{align}
\mu_i&=\exdi{\ve{Z};\ve{\theta}_i}{T(\ve{Z})},\\
\sigma_i&=\sqrt{ \exdi{\ve{Z};\ve{\theta}_i}{\big(T(\ve{Z})- \mu_i\big)^2} },
\end{align}
the asymptotic performance is then given by
\begin{align}
P_{\text{D}} &= \Prob{T(\ve{Z})>\xi | \mathcal{H}_1}\overset{a}{=} \qfunc{\frac{\xi-\mu_1}{\sigma_1}},\\
P_{\text{FA}}&= \Prob{T(\ve{Z})>\xi | \mathcal{H}_0}\overset{a}{=} \qfunc{\frac{\xi-\mu_0}{\sigma_0}},
\end{align}
where $Q(u)$ denotes the Q-function. Consequently, for a desired $P_{\text{FA}}$, the decision threshold is 
\begin{align}
\xi\overset{a}{=}\qfuncinv{P_{\text{FA}}}{\sigma_0}+\mu_0,
\end{align}
resulting in the asymptotic probability of detection
\begin{align}\label{def:probability:detection:deflection}
P_{\text{D}}(P_{\text{FA}})&\overset{a}{=}\qfunc{ \qfuncinv{P_{\text{FA}}} \frac{{\sigma_0}}{{\sigma_1}} - \frac{\mu_1 - \mu_0}{{\sigma_1}} }.
\end{align}
Writing
\begin{align}\label{natural:diff}
\ve{b}&=\ve{\beta}(\ve{\theta}_1)- \ve{\beta}(\ve{\theta}_0),
\end{align}
the log-likelihood ratio for exponential family models is
\begin{align}
\ln L(\ve{Z}) = \sum_{k=1}^{K} \ve{b}^{\T}  \ve{\phi}(\ve{z}_k) - K(\lambda(\ve{\theta}_1) - \lambda(\ve{\theta}_0)),
\end{align}
such that with the empirical mean of the sufficient statistics
\begin{align}
\ve{\bar{\phi}}&= \frac{1}{K} \sum_{k=1}^{K} \ve{\phi}(\ve{z}_k),
\end{align}
a likelihood-based test statistic is
\begin{align}\label{def:test:statistic}
T(\ve{Z})&= \ve{b}^{\T} \ve{\bar{\phi}}.
\end{align}
The mean and standard deviation of the test are
\begin{align}
\mu_i=\ve{b}^{\T} \ve{\mu}_{\ve{\phi}}(\ve{\theta}_i),\quad \sigma_i=\sqrt{\frac{1}{K}\ve{b}^{\T}  \ve{R}_{\ve{\phi}}(\ve{\theta}_i) \ve{b}},
\end{align}
where
\begin{align}\label{replacement:exp:family:mean}
\ve{\mu}_{\ve{\phi}}(\ve{\theta})&=\exdi{\ve{{z}};\ve{\theta}}{  \ve{\phi}(\ve{z}) },\\
\label{replacement:exp:family:variance}
\ve{R}_{\ve{\phi}}(\ve{\theta})&=\exdi{\ve{{z}};\ve{\theta}}{\ve{\phi}(\ve{z})\ve{\phi}^{\T}(\ve{z})}-\ve{\mu}_{\ve{\phi}}(\ve{\theta})\ve{\mu}_{\ve{\phi}}^{\T}(\ve{\theta}).
\end{align}
For Gaussian models \eqref{multivariate:gauss},
\begin{align}
\ve{\beta}(\ve{\theta})&=-\frac{1}{2}\vec{\ve{R}_{\ve{y}}^{-1}(\ve{\theta})},\\
\ve{\phi}(\ve{y})&=\vec{\ve{y}\ve{y}^{\T}},\\
\ve{\mu}_{\ve{\phi}}(\ve{\theta})&=\vec{\ve{R}_{\ve{y}}(\ve{\theta})},
\end{align}
and the matrix \eqref{replacement:exp:family:variance} can be determined through Isserlis' theorem. If the model \eqref{replacement:exp:family} is unspecified, a favorable choice of $\ve{\phi}(\ve{z})$ and $\ve{b}$ has to be found. For the $1$-bit analysis, we use
\begin{align}\label{aux:statistics:quantizer}
\ve{\phi}(\ve{z})&=\ve{\Phi}\vec{\ve{z} \ve{z}^{\T}},
\end{align}
where the matrix $\ve{\Phi}$ eliminates duplicate and diagonal statistics. This is potentially suboptimal as \eqref{aux:statistics:quantizer} does, in general, not contain all sufficient statistics of a multivariate Bernoulli distribution. The missing statistics are absorbed in the carrier of \eqref{replacement:exp:family} and, therefore, do not contribute to the decision process. For the natural parameter difference \eqref{natural:diff}, we use
\begin{align}\label{aux:nat:diff:quantizer}
\ve{b}= \ve{R}_{\ve{\phi}}^{-1}(\ve{\theta}_1) \ve{\mu}_{\ve{\phi}}(\ve{\theta}_1) - \ve{R}_{\ve{\phi}}^{-1}(\ve{\theta}_0) \ve{\mu}_{\ve{\phi}}(\ve{\theta}_0)
\end{align}
as it maximizes the distance of the asymptotic outcome in \eqref{def:test:statistic} under both hypotheses. The mean of the statistics \eqref{aux:statistics:quantizer},
\begin{align}
\ve{\mu}_{\ve{\phi}}(\ve{\theta})&=\exdi{\ve{z};\ve{\theta}}{\ve{\phi}(\ve{z})}= \ve{\Phi}\vec{\ve{R}_{\ve{z}}(\ve{\theta})},
\end{align}
is obtained by the arcsine law \cite[pp. 284]{Thomas69},
\begin{align}\label{covariance:quantized}
\ve{R}_{\ve{z}}(\ve{\theta})&=\frac{2}{\pi} \arcsin{ \ve{\Sigma}_{\ve{y}}(\ve{\theta}) },\notag\\
\ve{\Sigma}_{\ve{y}}(\ve{\theta})&=\diag{\ve{R}_{\ve{y}}(\ve{\theta})}^{-\frac{1}{2}} \ve{R}_{\ve{y}}(\ve{\theta}) \diag{\ve{R}_{\ve{y}}(\ve{\theta})}^{-\frac{1}{2}}.
\end{align}
Further, the evaluation of \eqref{replacement:exp:family:variance} requires determining the matrix
\begin{align}
\ve{C}(\ve{\theta})=\exdi{\ve{z};\ve{\theta}}{\vec{\ve{z} \ve{z}^{\T}}\vec{\ve{z} \ve{z}^{\T}}^{\T}}
\end{align}
which is possible \cite{SteinWSA16} by the arcsine law and the orthant probability of the quadrivariate Gaussian distribution \cite{Sinn11}.
\section{Results}\label{sec:results}
We apply the results to discuss the design of a $1$-bit array GPS spectrum monitoring system. The task is to check a spectral band with two-sided bandwidth $B=2.046$ Mhz, centered at $1.57$ GHz, for a source signal from a specific direction. The receiver samples the low-pass filtered receive signal at Nyquist rate, i.e., $f_s=B$, such that $K=2046$ array snapshots are available within a millisecond.
\pgfplotsset{legend style={rounded corners=2pt,nodes=right}}
\begin{figure}[!h]
\centering
\begin{tikzpicture}[scale=1]

  	\begin{axis}[ylabel=$\chi\text{ [dB]}$,
  			xlabel=$\text{Number of Sensors } S$,
			grid,
			ymin=-9.0,
			ymax=1.0,
			xmin=2,
			xmax=20,
			legend pos=south east]
			
			\addplot[green, style=solid, line width=0.75pt,smooth,every mark/.append style={solid}, mark=star, mark repeat=1] table[x index=0, y index=2]{DetectionPerformance_NumberOfSensors_SNR-15_OBS2046_zeta45.txt};
			\addlegendentry{$\text{SNR}=-15$ dB}

			\addplot[red, style=solid, line width=0.75pt,smooth,every mark/.append style={solid}, mark=square*, mark repeat=1] table[x index=0, y index=2]{DetectionPerformance_NumberOfSensors_SNR-18_OBS2046_zeta45.txt};
			\addlegendentry{$\text{SNR}=-18$ dB}
			
			\addplot[blue, style=solid, line width=0.75pt,smooth,every mark/.append style={solid}, mark=diamond*, mark repeat=1] table[x index=0, y index=2]{DetectionPerformance_NumberOfSensors_SNR-21_OBS2046_zeta45.txt};
			\addlegendentry{$\text{SNR}=-21$ dB}
			
			\addplot[black, style=solid, line width=0.75pt,smooth,every mark/.append style={solid}, mark=otimes*, mark repeat=1] table[x index=0, y index=2]{DetectionPerformance_NumberOfSensors_SNR-24_OBS2046_zeta45.txt};
			\addlegendentry{$\text{SNR}=-24$ dB}
			
	\end{axis}
	
\end{tikzpicture}
\caption{Quality vs. Array Size ($K=2046, \zeta=45^{\circ}$)}
\label{fig:roc:quality:sensors}
\end{figure}
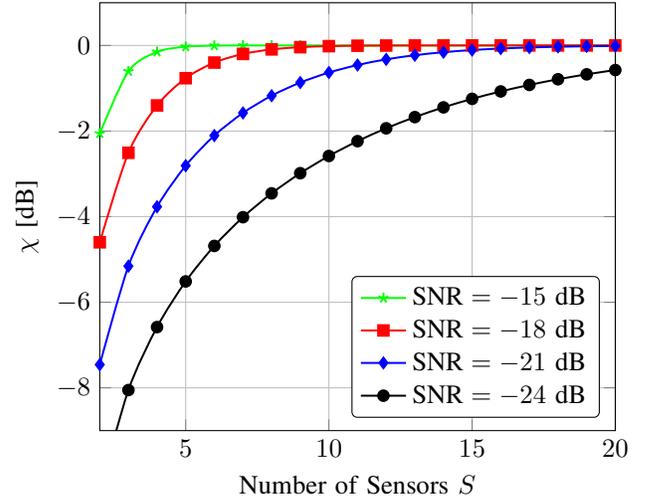
The upper triangular area under the receiver operating characteristic (ROC) curve,
\begin{align}\label{roc:quality}
\chi=2\int_{0}^{1}P_{\text{D}}(u) \rm{d}u-1,
\end{align}
determines the system quality regarding the detection task. For different signal-to-noise ratios (SNR), $\gamma_0=0$ vs. $\gamma_1=\sqrt{\text{SNR}}$, Fig. \ref{fig:roc:quality:sensors} shows the $1$-bit system quality for an exemplary setting with $K=2046, \zeta=45^{\circ},$ versus the number of array elements $S$. While for very weak sources ($\text{SNR}=-24$ dB) more than $S=20$ sensors are required to provide high performance, at a power level of $\text{SNR}=-15$ dB already $S=5$ antennas suffice to operate close to a perfect monitoring system with $\chi=1$.
\begin{figure}[!h]
\centering
\begin{tikzpicture}[scale=1]

  	\begin{axis}[ylabel=$\chi\text{ [dB]}$,
  			xlabel=$\text{Observation Time [ms]}$,
			grid,
			ymin=-7.0,
			ymax=1.0,
			xmin=0.1,
			xmax=10,
			legend pos=south east]
			
			\addplot[green, style=solid, line width=0.75pt,smooth,every mark/.append style={solid}, mark=star, mark repeat=10] table[x index=0, y index=2]{DetectionPerformance_NumberOfSamples_SNR-15_S8_zeta30.txt};
			\addlegendentry{$\text{SNR}=-15$ dB}

			\addplot[red, style=solid, line width=0.75pt,smooth,every mark/.append style={solid}, mark=square*, mark repeat=20] table[x index=0, y index=2]{DetectionPerformance_NumberOfSamples_SNR-18_S8_zeta30.txt};
			\addlegendentry{$\text{SNR}=-18$ dB}
			
			\addplot[blue, style=solid, line width=0.75pt,smooth,every mark/.append style={solid}, mark=diamond*, mark repeat=20] table[x index=0, y index=2]{DetectionPerformance_NumberOfSamples_SNR-21_S8_zeta30.txt};
			\addlegendentry{$\text{SNR}=-21$ dB}
			
			\addplot[black, style=solid, line width=0.75pt,smooth,every mark/.append style={solid}, mark=otimes*, mark repeat=20] table[x index=0, y index=2]{DetectionPerformance_NumberOfSamples_SNR-24_S8_zeta30.txt};
			\addlegendentry{$\text{SNR}=-24$ dB}
			
	\end{axis}
	
\end{tikzpicture}
\caption{Quality vs. Observation Time ($S=8,\zeta=30^{\circ}$)}
\label{fig:roc:quality:time}
\end{figure}
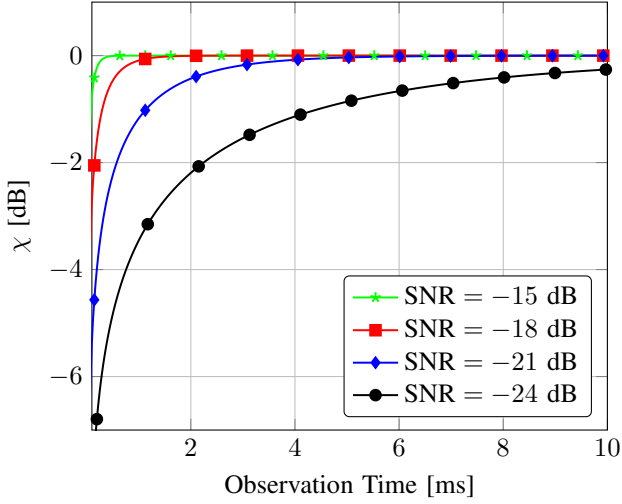
To determine a favorable observation length, for $S=8,\zeta=30^{\circ},$ Fig. \ref{fig:roc:quality:time} shows $\chi$ for different numbers of samples. While reliable detection with $\text{SNR}=-24$ dB requires sampling more than $10$ ms, the decision at $\text{SNR}=-15$ dB can be made trustworthy within less than $1$ ms. Fig. \ref{fig:simulated:performance} depicts the analytic and simulated performance (using $10^5$ realizations) with $S=8, K=100, \zeta=15^{\circ}$. At moderate sample size, the asymptotic results already show a good correspondence with the Monte-Carlo simulations.
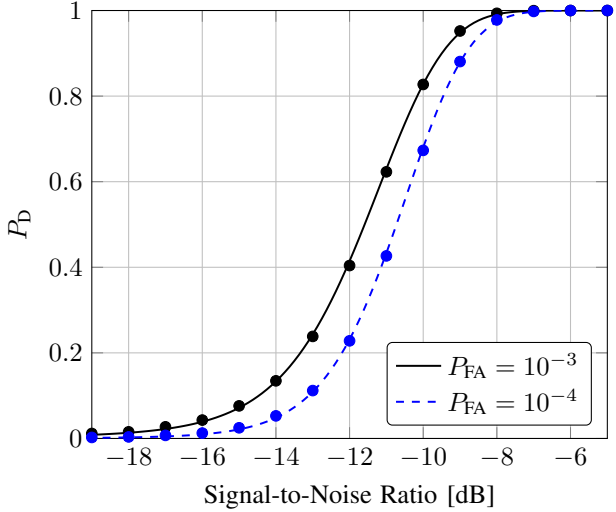
\begin{figure}[!h]
\centering
\begin{tikzpicture}[scale=1]

  	\begin{axis}[ylabel=$P_{\text{D}}$,
  			xlabel=$\text{Signal-to-Noise Ratio [dB]}$,
			grid,
			ymin=0,
			ymax=1.0,
			xmin=-19,
			xmax=-5,
			legend pos=south east]
			
			\addplot[black, style=solid, line width=0.75pt,smooth] table[x index=0, y index=2]{DetectionPerformance_Analytic_pfa0.001_S8_OBS100_zeta15.txt};
			\addlegendentry{$P_{\text{FA}}=10^{-3}$}
			
			\addplot[blue, style=dashed, line width=0.75pt,smooth] table[x index=0, y index=2]{DetectionPerformance_Analytic_pfa0.0001_S8_OBS100_zeta15.txt};
			\addlegendentry{$P_{\text{FA}}=10^{-4}$}

			\addplot[black,only marks , every mark/.append style={solid}, mark=otimes*, mark repeat=1] table[x index=0, y index=3]{DetectionPerformance_Simulation_pfa0.001_S8_OBS100_zeta15.txt};

		        \addplot[blue,only marks , every mark/.append style={solid}, mark=otimes*, mark repeat=1] table[x index=0, y index=3]{DetectionPerformance_Simulation_pfa0.0001_S8_OBS100_zeta15.txt};
			
	\end{axis}
	
\end{tikzpicture}
\caption{Analysis vs. Simulation ($S=8, K=100, \zeta=15^{\circ}$)}
\label{fig:simulated:performance}
\end{figure}
\section{Conclusion}
\label{sec:conclusion}
We have derived achievable detection rates with a large number of radio measurements obtained with a low-complexity sensor array performing $1$-bit analog-to-digital conversion. Discussing the simple binary hypothesis test in the framework of the exponential family enables circumventing the intractability of the $1$-bit array likelihood. Its difficult characterization forms a fundamental obstacle to the application of analytic tools in statistical signal and information processing to problems with coarsely quantized multivariate data. Using the analytic results to determine the spectrum monitoring capability with wireless multi-antenna receivers shows that, under the right system design, radio frequency measurements from binary arrays are sufficient to perform the processing task of signal detection in a fast and reliable way.

\end{document}